\documentclass[times,twocolumn]{article}
\usepackage[utf8]{inputenc}
\usepackage{amsmath}
\usepackage[margin=0.5in]{geometry}
\usepackage{graphicx}
\usepackage[english]{babel,varioref}
\usepackage[font=small, labelfont=bf]{caption}
\usepackage{subcaption}
\usepackage{multimedia}
\usepackage{import}
\usepackage{multicol}
\usepackage{amsfonts}
\usepackage{listings}
\usepackage{color}
\usepackage{tabularx}

\usepackage[
  maxcitenames=2,
  doi=false,
  backend=biber,
  firstinits=true,
  style=numeric-comp,
  sorting=none,]
{biblatex}
\usepackage{tikz}
\usetikzlibrary{external,positioning} 
\usepackage{siunitx}

\definecolor{urxvtstring}{HTML}{FFA07A}
\definecolor{urxvtkeyword}{HTML}{00FFFF}
\definecolor{urxvtcomment}{HTML}{FF7F24}
\definecolor{urxvtemph}{HTML}{98fb98}
\definecolor{urxvtgrey}{HTML}{93A1A1}
\lstMakeShortInline[columns=fixed]|
\lstdefinestyle{mystyle}{
  commentstyle=\color{urxvtcomment},
  keywordstyle=\color{urxvtkeyword},
  emph={post,__init__},
  emphstyle=\color{urxvtemph},
  numberstyle=\tiny\color{urxvtgrey},
  stringstyle=\color{urxvtstring},
  basicstyle=\ttfamily\footnotesize,
  breakatwhitespace=false,
  breaklines=true,
  captionpos=b,
  keepspaces=true,
  numbers=left,
  xleftmargin=10pt,
  numbersep=5pt,
  showspaces=false,
  showstringspaces=false,
  showtabs=false,
  tabsize=2,
  texcl=false, 
  columns=flexible,
}
\definecolor{lessinteresting}{gray}{0.5}

\newcommand{\gray}[1]{\textcolor{gray}{#1}}

\author{
Erik Wallin\textsuperscript{1}\thanks{Corresponding author: erik.wallin@umu.se},
\\
\textsuperscript{1}Department of Physics, Umeå University
}

\title{A modular and extensible library for parameterized terrain generation}

\begin{document}
\maketitle


\begin{abstract}
Simulation-driven development of intelligent machines benefits from artificial terrains with controllable, well-defined characteristics.
However, most existing tools for terrain generation focus on artist-driven workflows and visual realism, with limited support for parameterization, reproducibility, or scripting.
We present a modular, Python-based library for procedural terrain generation that enables users to construct complex, parameterized terrains by chaining together simple modules.
The system supports both structured and noise-based terrain elements, and integrates with Blender for rendering and object placement.
The framework is designed to support applications such as generating synthetic terrains for training machine learning models or producing ground truth for perception tasks.
By using a minimal but extensible set of modules, the system achieves high flexibility while remaining easy to configure and expand.
We demonstrate that this enables fine-grained control over features such as slope, roughness, and the number of rocks, as well as extension to additional measures.
This makes it well suited for workflows that demand reproducibility, variation, and integration with automated pipelines.
\end{abstract}

\providecommand{\keywords}[1]
{
  \small
  \textbf{\textit{Keywords---}} #1
}

\section{Introduction}
To develop the next generation of intelligent machines, simulations play a crucial role, and for many applications these must include artificial terrains with controllable, well-defined characteristics.
This is particularly important for training trafficability models for forestry vehicles \cite{wallin2022learning}, designing and testing new machines, and developing methods for the automatic estimation of terrain features from scanned forest data.

Procedural generation of environments has been explored for over four decades, particularly within the domains of computer graphics and video game development.
In these contexts, terrain generation is one component among others, such as the procedural creation of human-made structures (e.g. buildings and roads) and other natural elements such as vegetation.
Common techniques for generating terrains include noise-based methods, utilizing Perlin or Simplex noise \cite{perlin1985image, perlin2002improving}.
\textcite{smelik2014survey} provides an overview of a wide range of procedural modelling techniques.
While these approaches are capable of producing a wide variety of landscapes, they often lack fine-grained control over specific terrain properties.
As a result, it can be difficult to ensure that the generated terrains satisfy particular characteristics.

In many applications, it is important to do precisely this: to analyze and control terrain properties across a wide range of scenarios.
This requires a parameterized generation process, where terrain characteristics can be specified explicitly and sampled from specified distributions.
To be practically useful, such a system must be powerful enough to produce complex and varied terrains, simple to configure, flexible enough to support diverse use cases, and extensible to accommodate future development.

Several tools and libraries for terrain generation exist, including artist-oriented software such as World Machine \cite{worldmachine} and Gaea \cite{gaea}, and game engine plugins like MapMagic2 \cite{mapmagic} and GAIA \cite{gaia}.
While these systems excel in visual design, they generally lack scriptability, modular composition, or explicit parameter control.
Moreover, they are often not well suited for simulation workflows requiring batch generation, reproducibility, or fine-grained variation in terrain characteristics.

Motivated by these needs, we present a flexible and extensible software library for generating artificial terrains with controllable characteristics:
\newline
\url{https://github.com/erikwallin86/artificial_terrains}

The main contributions of this work are:
\begin{enumerate}
\item a modular, Python-based terrain generation library that supports both structured and noise-based terrain elements, with integrated rendering and object placement via Blender,
\item a set of methods for parametrizing terrains using metrics such as surface roughness, slope, and rock number density,
\item examples demonstrating how modular components can be composed to generate complex terrain configurations with controllable features.
\end{enumerate}
To further demonstrate the capabilities and structure of the system, the full set of commands used to generate all terrains, analysis, and figures in this paper is provided in Appendix~\ref{subsec:commands_for_figures}.

The remainder of the paper is organized as follows. In Section~\ref{sec:method}, we present the structure and functionality of the library.
Section~\ref{sec:results} shows how terrains can be parameterized, in a manner that is extensible to new measures.
We conclude with a discussion and final remarks in Section~\ref{sec:discussion} and Section~\ref{sec:conclusion}.

\section{Method}
\label{sec:method}
We generate terrains using a modular system, with a list of \emph{modules} executed in order.
Each module can take input and generate output, and the output of one module is passed as input to the next, as if connected through a \emph{pipe}.
This interface is in part inspired by UNIX, where relatively simple programs with specific tasks can be combined via piping to accomplish rather complex operations.
Another source of inspiration for this setup is ImageMagick, a command-line image processing program, powerful enough to do ``almost anything''.
Terrains represented as 2D regular grid heightmaps are very similar to (grayscale) images, and we also want to be able to generate ``almost any terrain''.
In ImageMagick, ``operators'' are defined and applied in order, and sub-results can be generating using parenthesis to group operations.
The program internally maintains a list of images, which can be manipulated with operations such as duplication, deletion, or reordering, and can be combined using a variety of compositing methods.

The idea is the same here --- to have a set of modules with relatively simple behaviours that are useful on their own, that can be combined to achieve arbitrarily complex results, and that can be easily extended with new modules.

To support this, we need a system for generating and combining terrains, ideally simple to understand, yet not limiting when it comes to achieving complex results.
Our setup involves managing terrains in two lists: \emph{temporary} and \emph{primary}.
The temporary list can be seen as a ``staging area'', where intermediate results are stored and prepared, before being combined to a more complex (sub)result.
This allows for keeping (sub)results in memory, while also preparing additional components.
\emph{Generating} modules create basic terrain elements which are added to the temporary list, where they can then be handled by a \emph{combining} module.
These combines lists of terrains using some mathematical operation, and place the resultant terrain in the primary list.
If the temporary list is empty, they operate directly on the primary list instead.

Terrains can be moved between lists, and operations can be applied to parts of lists.
\emph{Modifier} modules can e.g. scale terrains by some factor.
\emph{Input/output} modules can save, load, or visualise generated terrains.

Necessary input to the generating modules is an \emph{extent} (m) and a \emph{grid size}, specifying the size, location and the number of raster points.
Default values are \verb|extent=[-25, 25, -25, 25]| and \verb|grid_size=[100, 100]|, but this can be modified using the modules \verb|Size|, \verb|Location|, \verb|Extent|, \verb|Resolution|, and \verb|GridSize|.

The following subsections cover the different types of modules and how they can be combined.
The system is built around a minimal set of general-purpose modules, designed to be combined flexibly, with the focus placed on developing this core set rather than maximizing the number of available modules.

\subsection{Settings}
To run the script, a \emph{save-dir} and a list of modules must be defined, and optionally \emph{settings} and/or a \emph{settings-file}.
\begin{lstlisting}[frame=single]
generate_terrain.py --save-dir folder [--settings-file ...] [--settings ... ]  --modules MODULES
\end{lstlisting}
Input to modules can be provided in different manners, besides being passed from one module to the next. Keyword arguments can be passed to a specific module, as a dict with in the format
\begin{lstlisting}[frame=single]
Module:"dict(parameter=value)"
\end{lstlisting}
or directly as the special \verb|default| keyword using
\begin{lstlisting}[frame=single]
Module:value
\end{lstlisting}
where \verb|value| can be an int, float, string, bool, or list.
Each module defines how the \verb|default| input is handled, depending on its specific functionality.
\emph{General settings} are passed as keyword arguments to all modules, and are defined as
\begin{lstlisting}[frame=single]
--settings parameter:value parameter2:value2
\end{lstlisting}
Settings can also be loaded from a yaml file, as
\begin{lstlisting}[frame=single]
--settings-file settings.yml
\end{lstlisting}
together with a settings file,
\begin{lstlisting}[frame=single, caption={Contents of \texttt{settings.yml}}]
# Example yaml settings file
parameter1 = value1
parameter2 = value2
\end{lstlisting}
Note that the list of modules itself can also be specified in the yaml file, such that it defines the entire setup.
Finally, parameters can be set by modules whose purpose is to set parameters for following modules, with one such (trivial) example being the \verb|Set| module,
\begin{lstlisting}[frame=single]
--modules Set:parameter=value Module
\end{lstlisting}
A more complex example may involve a module that samples parameters from either a probability distribution or a deterministic sequence, influencing multiple downstream modules to generate terrains with desired properties.

When a setting is defined in multiple places, the following order of precedence applies (with later items overriding earlier ones): settings-file $\rightarrow$ general settings $\rightarrow$ output from previous modules $\rightarrow$ specific settings.

\subsection{Generating modules}
\label{subsec:label}
Generating modules are those that produce one or several basic terrain elements.
There are currently two types of generating modules: those that generate terrain elements from random noise, and those that generate terrain elements from functions.
We begin by describing the noise-based modules, and will return later to the function-based ones --- including how they can be combined with lists of (random) obstacles to also produce varied and realistic terrain elements.

\begin{figure}[ht]
  \centering
  \includegraphics[width=0.85\columnwidth]{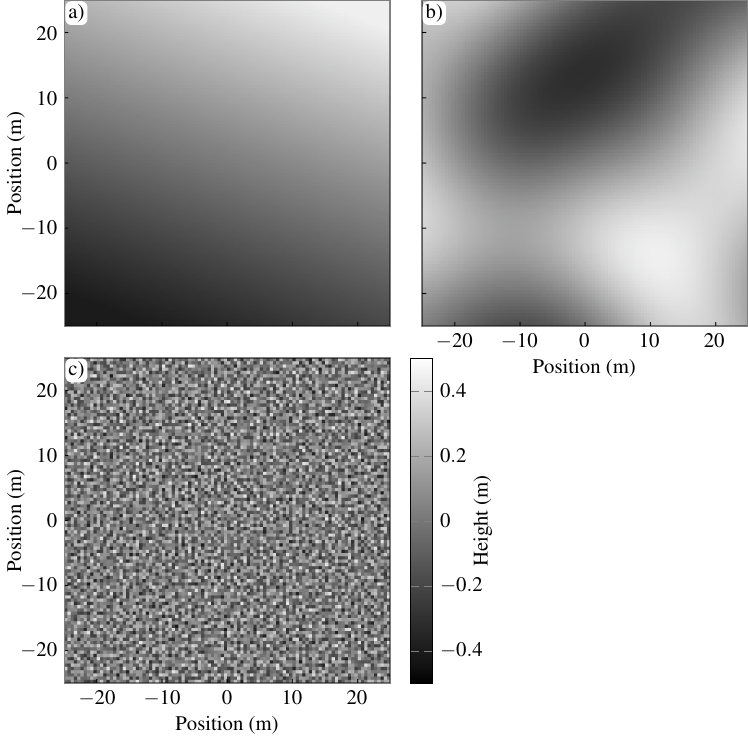}
  \caption{\label{fig:basic} Plots of the default output from \texttt{Basic}.}
\end{figure}

\begin{figure}[ht]
  \centering
  \includegraphics[width=0.85\columnwidth]{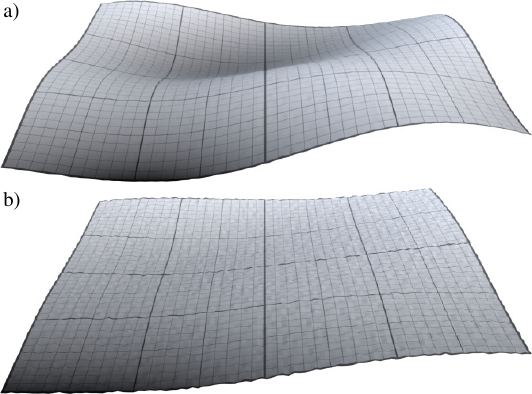}
  \caption{\label{fig:basic_combined} Two different weighted sums of the terrain elements in Fig.~\ref{fig:basic}, with weights \texttt{[5, 8, 0.1]} and \texttt{[10,3,0.3]}.}
\end{figure}

\begin{figure}[ht]
  \centering
  \includegraphics[width=0.85\columnwidth]{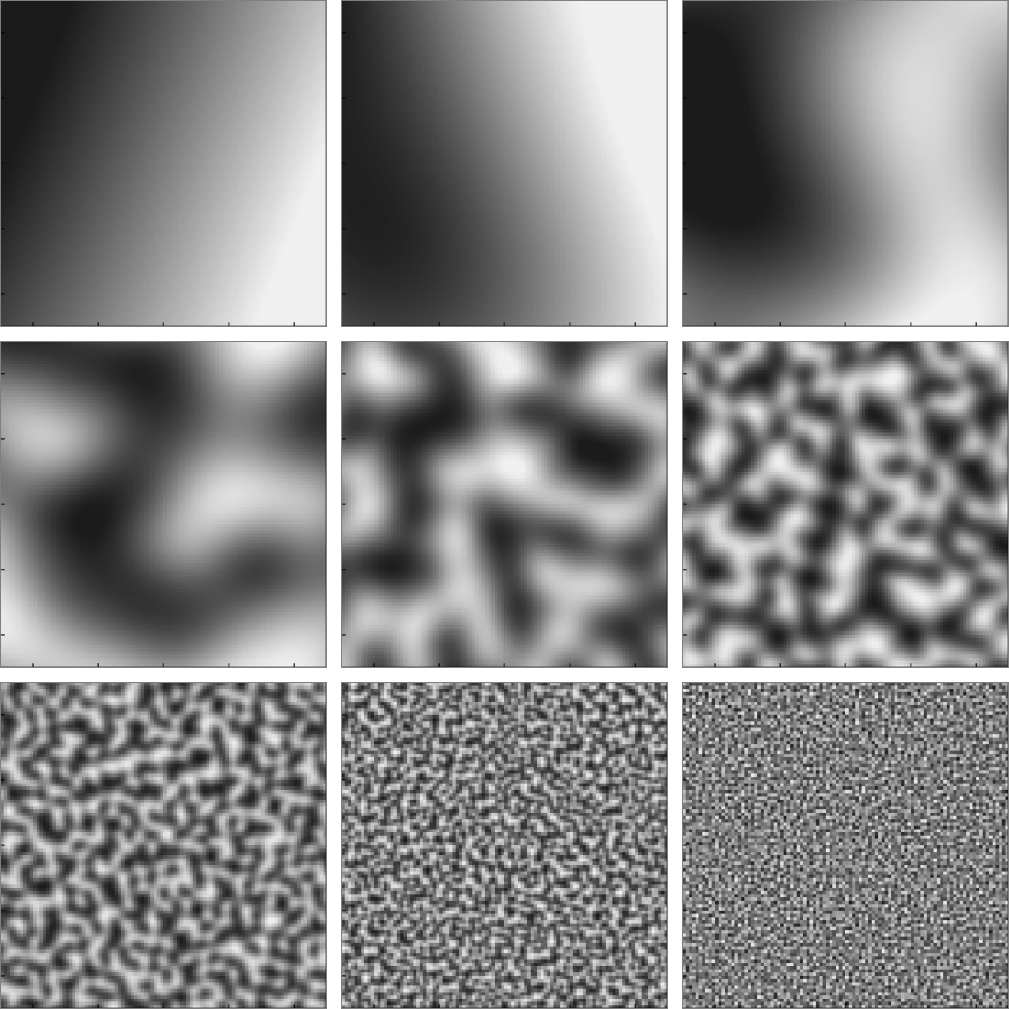}
  \caption{\label{fig:octaves} Plots of the default output from \texttt{Octaves}, each with size $50 \times 50$~m and with $100\times100$ grid points.}
\end{figure}

\begin{figure}[ht]
  \centering
  \includegraphics[width=0.85\columnwidth]{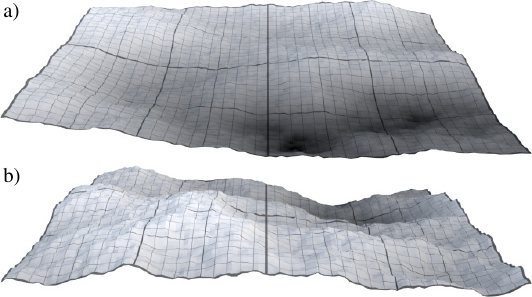}
  \caption{\label{fig:octaves_combined} Two examples of weighted sums of the terrain elements in Fig.~\ref{fig:octaves}.}
\end{figure}

We utilize Simplex noise to generate terrain elements of different \emph{length scales}, applying a scale factor as a function of the length scale so that the resulting amplitude is approximately $0.5$.
The module \verb|Basic| by default generates 3 terrains elements, with length scales in metres given by \verb|scale=[400, 32, 0.5]|.
The resultant terrain elements, from running
\begin{lstlisting}[frame=single, escapechar={|}]
|\gray{python}| |\gray{generate\_terrain.py}| |\gray{--save-dir}| |\gray{folder}| |\gray{--modules}| Basic Plot
\end{lstlisting}
can be seen in Fig.~\ref{fig:basic}, composited from the output of the \texttt{Plot} module.
The terrain elements can be combined into a single terrain using a weighted sum, where the choice of weights naturally has a significant impact on the structure of the resulting terrain, as can be seen from the two examples in Fig.~\ref{fig:basic_combined}.
The number of generated terrain elements and their length scales can be defined in a list, or with a single value interpreted as a list of length 1,
\begin{lstlisting}[frame=single, escapechar={|}]
Basic:[100,20,5,1]
Basic:100
\end{lstlisting}

The module \texttt{Octaves} generates a number of terrain elements, where the length scale is halved for each subsequent element, starting from some defined initial value.
It also creates a list of amplitudes, based on a starting value that is scaled by a \texttt{persistence} factor at each step.
The scaling is not directly applied to the terrain elements; instead, the values are passed along as a \texttt{weights} parameter, and can be used by a subsequent \texttt{WeightedSum} module.
We allow for scaling each amplitude by a factor $\sim \mathcal{N}(1, \texttt{random\_amp})$ as well as assigning a \texttt{random\_sign}, possibly allowing for great variation from the same set of terrain elements.
The output using the default parameters, with an initial length scale of $128$~m, initial amplitude of $10$~m, persistence $0.60$, and \texttt{random\_amp} 0.5 can be seen in Fig.~\ref{fig:octaves}.
Two examples of the subsequent weighted sums can be seen in Fig.~\ref{fig:octaves_combined}.

\begin{figure}[ht]
  \centering
  \includegraphics[width=0.85\columnwidth]{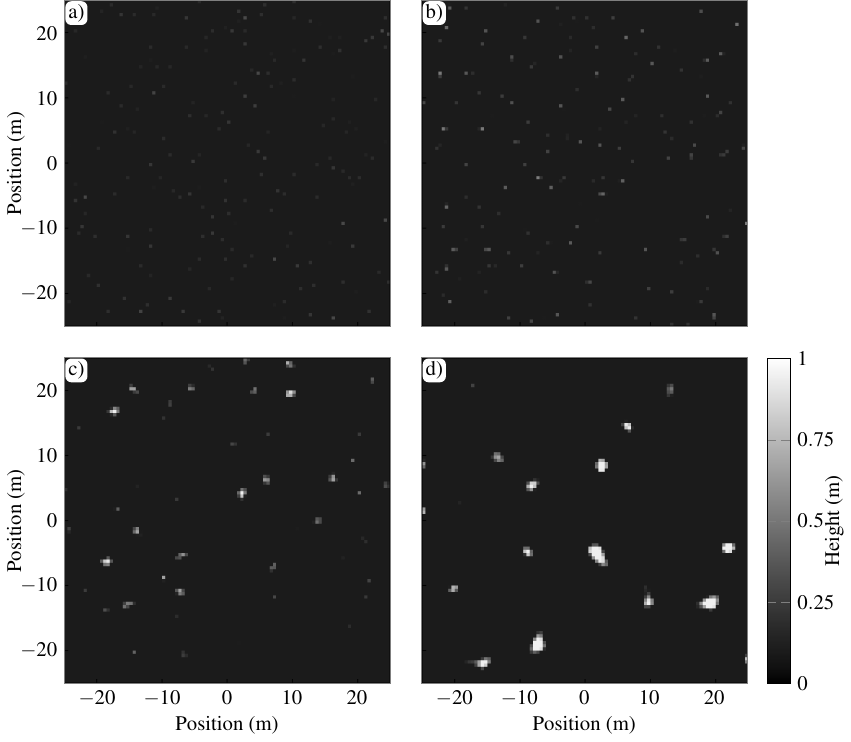}
  \caption{\label{fig:rocks} Plots of the default output from \texttt{Rocks}}
\end{figure}

\begin{figure}[ht]
  \centering
  \includegraphics[width=0.85\columnwidth]{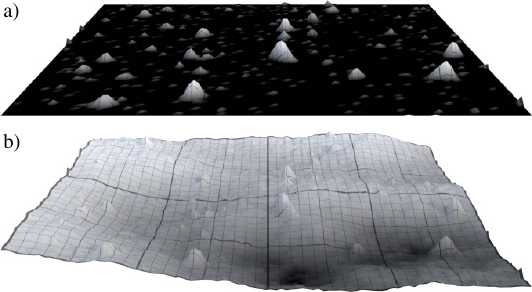}
  \caption{\label{fig:rocks_combined} a) Terrain elements from Fig.~\ref{fig:rocks}, combined by simple addition. b) The same resultant rocks, added to the terrain of Fig.~\ref{fig:octaves_combined}a.}
\end{figure}

The modules \texttt{Rocks} and \texttt{Holes} are designed to produce rock-like and hole-like structures from noise.
They generate simplex noise terrain elements of different length scales, and \emph{clip} these at some \texttt{fraction=0.8} of the maximum/minimum value, such that only the top/bottom are used.
The \texttt{height} can be defined in a list, but is by default half the length scale for rocks, and one fourth for holes.
Output from the \texttt{Rocks} module can be seen in Fig.~\ref{fig:rocks}, with the result of combining these using addition seen in Fig.~\ref{fig:rocks_combined}.

\begin{figure}[ht]
  \centering
  \includegraphics[width=0.85\columnwidth]{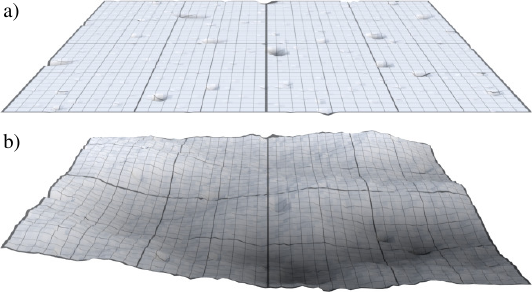}
  \caption{\label{fig:holes_combined} a) \texttt{Holes} module output, combined by simple addition. b) The same resultant holes, added to the terrain of Fig.~\ref{fig:octaves_combined}a.}
\end{figure}

Output from the \texttt{Holes} module, combined using addition, can be seen in Fig.~\ref{fig:holes_combined}.

\subsection{Combining modules}
\label{subsec:label}
The \emph{combining} modules combine multiple terrain elements to one, by means of some mathematical operation.
These are currently the \texttt{Combine} and the \texttt{WeightedSum} modules, where the latter is the most straight forward as it simply performs a weighted sum, given the list of terrains and a list of \texttt{weights}.
\texttt{Combine} combines a list of terrains to one, using either of the operations \texttt{Add}, \texttt{Prod}, \texttt{Min}, or \texttt{Max}.
It operates primarily on the \emph{temporary} list and secondary on the \emph{primary}, and by default consumes all entries in the list.
However, it takes an optional input \texttt{last}, an integer specifying how many of the last terrains to consume.

\subsection{Obstacles and generating functions}
Here we present modules that generate terrain elements from functions, including how many of them can be combined with lists of (random) \emph{obstacles} parameters in order to produce varied terrain elements.
The one exception to the latter is the \texttt{Function} module, which generates a single terrain element from an expression, and which we will describe later on.

\begin{figure}[ht]
  \centering
  \includegraphics[width=0.85\columnwidth]{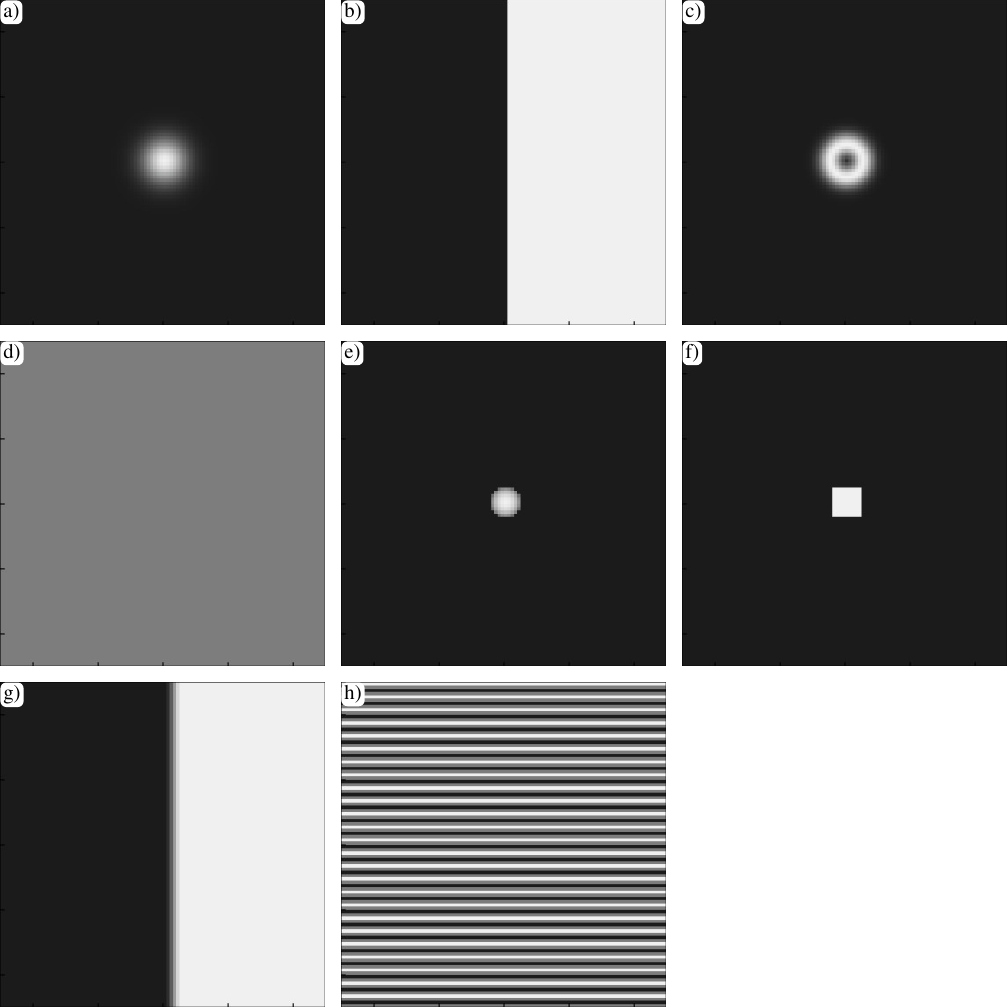}
  \caption{\label{fig:generating_function} Plots of the default output from the 8 generating functions, each with size $50 \times 50$~m and with $100\times100$ grid points.}
\end{figure}

The current list of generating function modules are \texttt{Gaussian}, \texttt{Step}, \texttt{Donut}, \texttt{Plane}, \texttt{Sphere}, \texttt{Cube}, \texttt{SmoothStep}, and \texttt{Sine}, where many of them are self-explanatory.
By default, each module produce a single terrain element. We can generate them all with the following command,
\begin{lstlisting}[frame=single, escapechar={|}]
|\gray{python}| |\gray{generate\_terrain.py}| |\gray{--save-dir}| |\gray{folder}| |\gray{--modules}| Gaussian Step Donut Plane Sphere Cube SmoothStep Sine Plot
\end{lstlisting}
and the resultant plots can be seen in Fig.~\ref{fig:generating_function}.
The generating functions share a common input interface, being defined by the parameters
\texttt{position}, \texttt{height}, \texttt{width}, \texttt{aspect}, \texttt{yaw\_deg}, and \texttt{pitch\_deg}, where each is a list of possible values.
Not all parameters are relevant for all functions, in which case they are ignored.
E.g. for \texttt{Plane}, which generates a flat, inclined plane, only \texttt{position}, \texttt{yaw\_deg} and \texttt{pitch\_deg} are of relevance. 

\begin{figure}[ht]
  \centering
  \includegraphics[width=0.85\columnwidth]{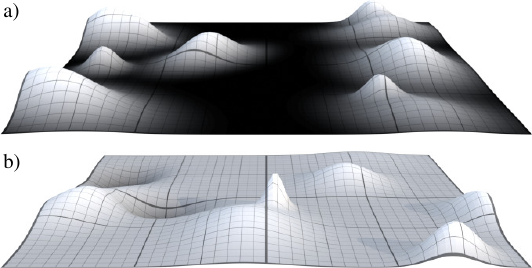}
  \caption{\label{fig:generating_function_2} a) Render of terrain with 10 Gaussian hills, b) the same but with the \texttt{height} distribution altered to also produce pits.}
\end{figure}

It is possible to specify lists of input, in order to generate multiple terrain elements from the modules above, e.g. generating Gaussian hills at specified locations.
However, often a more convenient workflow is to generate these lists using another module.
\texttt{Random} is such an example, and generates list entries for the different parameters by sampling from specified \emph{Distributions}.
In the following command, lists with 10 entries are randomly sampled, such that the \texttt{Gaussian} module produce 10 terrain elements.
These are then combined by addition, with the resultant terrain seen in Fig.~\ref{fig:generating_function_2}a.
\begin{lstlisting}[frame=single, escapechar={|}]
|\gray{python}| |\gray{generate\_terrain.py}| |\gray{--save-dir}| |\gray{folder}| |\gray{--modules}| Random:10 Gaussian Combine:Add Plot
\end{lstlisting}

By default, the \texttt{Random} module samples parameters from uniform distributions, but normal, exponential, and beta distributions are also supported.
Alternatively, deterministic sequences such as \texttt{arange}, \texttt{linspace}, and \texttt{logspace} can be used.
All parameters have fixed default distributions, except for \texttt{position}, where the x- and y-components are uniformly sampled with limits provided by the \texttt{extent}.
The distributions can be altered using the \texttt{SetDistribution} module, with the syntax \texttt{SetDistribution:"parameter=distribution(arguments)"}.
In some interpreters, quotation marks may be required around parentheses; to avoid this, square brackets can be used instead.
In the following example we alter the range of the uniform sampling to also produce Gaussian pits using $\texttt{height} \sim \mathcal{U}(-5, 5)$
\begin{lstlisting}[frame=single, escapechar={|}]
|\gray{python}| |\gray{generate\_terrain.py}| |\gray{--save-dir}| |\gray{folder}| |\gray{--modules}| SetDistribution:height=uniform[-5,5] Random:10 Gaussian Combine:Add Plot
\end{lstlisting}
with the resultant terrain seen in Fig.~\ref{fig:generating_function_2}b.

\begin{figure}[ht]
  \centering
  \includegraphics[width=0.85\columnwidth]{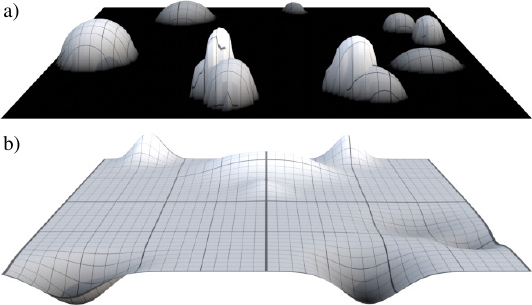}
  \caption{\label{fig:distributions} a) Render of terrain with 10 spheres, with positions sampled from a \texttt{Donut}, b) Render of terrain with 10 Gaussian shapes, with heights taken from lookup from an angled plane.}
\end{figure}

Besides the above method of sampling parameters from 1D distributions, there are two other means of changing the distributions from which points are sampled.
Firstly, \texttt{position} can be sampled from a \emph{2D distribution function}, where a terrain can be interpreted as the distribution function.
The \texttt{AsProbability} module take the \emph{most recent} terrain (i.e. last terrain in the temporary list), clips it at $0$ to remove negative values, and normalise it to get a probability $P(x, y)$.
Positions are then sampled from this distribution,
\begin{equation}
  \label{}
p_i=(x_i, y_i) \sim P(x, y).
\end{equation}
In the following example, a 40 m donut is used as a probability distribution, with the effect of the following 10 spheres of varying size and height being sampled along a circle,
\begin{lstlisting}[frame=single, escapechar={|}]
|\gray{python}| |\gray{generate\_terrain.py}| |\gray{--save-dir}| |\gray{folder}| |\gray{--modules}| Donut:"dict(width=40)" AsProbability Random:10 Sphere Combine
\end{lstlisting}
as can be seen in Fig.~\ref{fig:distributions}a.

Secondly, given a set of positions $\{p_i\}_{i=1}^N$, additional parameters can be sampled from 2D data as well, with a terrain used as a \emph{lookup function} $\Phi(p_i)$,
\begin{equation}
  \label{}
  \phi_i = \Phi(p_i) = \Phi(x_i, y_i).
\end{equation}
where $\phi$ represent some parameter.
The \texttt{AsLookupFor} module take the \emph{most recent} terrain (i.e. last terrain in the temporary list), and assign it as a lookup function for a given parameter.
The syntax can be seen in the following example, where we utilize an angled plane as lookup for the height parameter, for 10 Gaussian hills,
\begin{lstlisting}[frame=single, escapechar={|}]
|\gray{python}| |\gray{generate\_terrain.py}| |\gray{--save-dir}| |\gray{folder}| |\gray{--modules}| Plane:"dict(pitch_deg=10)" AsLookupFor:height Random:10 Gaussian Combine
\end{lstlisting}
with the heights increasing linearly from approximately $-4$~m to $4$~m along the y-direction, as seen in Fig.~\ref{fig:distributions}b.

\begin{figure}[ht]
  \centering
  \includegraphics[width=0.6\columnwidth]{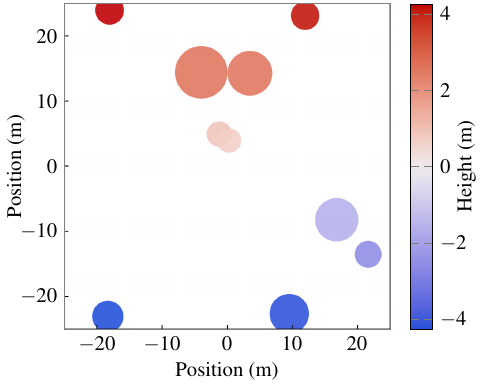}
  \caption{\label{fig:obstacles} Plot showing \texttt{position}, \texttt{width}, and \texttt{height} of 10 obstacles, with \texttt{aspect}, \texttt{yaw\_deg}, \texttt{pitch\_deg} not visualised. The set of obstacles is the same as in Fig.~\ref{fig:distributions}b, i.e. sampled with heights taken from a lookup function given by an angled plane.}
\end{figure}

The set of parameters \texttt{position}, \texttt{height}, \texttt{width}, \texttt{aspect}, \texttt{yaw\_deg}, \texttt{pitch\_deg} is not only useful for generating terrain elements using the described functions, but can also more generally describe local, scattered structures within the environment.
For that reason, we refer to these as \emph{obstacles}.
Obstacles lists can be saved/loaded to/from files using \texttt{SaveObstacles}/\texttt{LoadObstacles}, stored either as arrays in NumPy \texttt{npz} files, or in a yaml-file.
They can also be visualised with a simple scatter-plot, using \texttt{PlotObstacles} as seen in Fig.~\ref{fig:obstacles}.

\begin{figure}[ht]
  \centering
  \includegraphics[width=0.85\columnwidth]{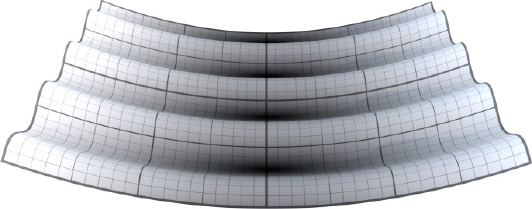}
  \caption{\label{fig:function} Render of terrain created with \texttt{Function} module.}
\end{figure}

Finally, there is one generating function that does not work with the obstacles framework.
The module \texttt{Function} instead generates a single terrain element from a given arithmetic expression involving the x- and y-coordinates.
The syntax can be seen in the following example,
\begin{lstlisting}[frame=single, escapechar={|}]
|\gray{python}| |\gray{generate\_terrain.py}| |\gray{--save-dir}| |\gray{folder}| |\gray{--modules}| Function:'5*(x/np.max(x))**2+np.sin(y/2)' Save
\end{lstlisting}
with a quadratic shape in the x-direction together with a sine shape in the y-direction, as seen in Fig.~\ref{fig:function}.
The expression can involve NumPy functions, and \texttt{r} can be used as a shorthand for a radial coordinate.
This is powerful not only because of the multitude of terrains one can generate, but also because of how it can be used for sampling and lookup using \texttt{AsProbability} and \texttt{AsLookupFor} modules, together with \texttt{Random}.

\subsection{Other modules}
We have covered modules for generating terrain elements, combining them, and plotting the output.
Here, we go over the remaining modules.
Some of these have already been used to generate the terrains presented in the figures, while others will be used later.
Modules related to \emph{Camera and Rendering} are saved for section~\ref{sec:blender},
and modules related to \emph{Control Flow} are saved for section \ref{subsec:multiple_terrains}.

The \emph{Input/Output} modules \texttt{Save} and \texttt{Load} can been used for saving and loading terrains.
There are a number of \emph{Modification} modules for altering terrain elements.
Most of these work on all terrains in the temporary or primary lists.
\texttt{Absolute} takes the absolute value ($\vert z_{xy}\vert$), \texttt{Smooth} applies Gaussian smoothing, and \texttt{Clip} clips values to a specified range.
\texttt{Negate} inverts terrain heights ($-z_{xy}$), \texttt{Add} adds a term ($a + z_{xy}$), and \texttt{Scale} adjust their magnitude ($a\times z_{xy}$ or $a_{xy}\times z_{xy}$).
\texttt{Scale} takes a (float) parameter \texttt{factor} as the default input, but a position-dependent factor $a_{xy}$ can be introduced by the module \texttt{AsFactor}, which then consumes the \emph{last} terrain.

\texttt{Print} and \texttt{DebugPlot} can be used to gain insight into the content of the pipe.

The \emph{Terrain analysis} modules \texttt{Roughness}, \texttt{Slope}, \texttt{SurfaceStructure} and \texttt{FindRocks} can be used to analyse generated terrains.
\texttt{FindRocks} identifies rocks using \texttt{scipy.ndimage.label} \cite{2020SciPy-NMeth} and returns an ``obstacles'' list containing \texttt{position}, \texttt{height}, and \texttt{width} (as well as dummy values for \texttt{aspect}, \texttt{yaw\_deg}, and \texttt{pitch\_deg}).

\texttt{LogData} and \texttt{LoadData} can be used for data processing.
\texttt{LogData} collects all non-terrain data from the pipe(s) to save as a NumPy \texttt{npz} archive when the script finishes.
When used in a loop, it stores values in lists and attempts to convert them to arrays.
If conversion using NumPy’s default constructor fails (e.g., for a list of lists with varying lengths), it instead attempts to \emph{concatenate} them.
\texttt{LoadData} loads an \texttt{npz} archive and can be used with \emph{Visualization} modules to plot data passed through the pipe.
\texttt{PlotScatter} creates scatter plots for all pairs of matching data, \texttt{PlotHistogram} generates histograms for each data entry, and \texttt{PlotLines} creates line plots for all data structured as \texttt{(lines $\times$ points)}.

Modules for \emph{Parametrisation} such as \texttt{SetRoughness} and \texttt{SetSlope} will be explored in section~\ref{subsec:parameterized_terrains}.

The entire list of modules can be seen in section~\ref{subsec:list_of_modules}.

\subsection{Blender}
\label{sec:blender}
The same script with the same module interface can be used together with Blender \cite{blender2025}, by calling it in the following way
\begin{lstlisting}[frame=single, escapechar={|}]
blender --python generate_terrain.py --
\end{lstlisting}
where \texttt{blender --python} is the Blender command to run a given python script, and \texttt{--} is an argument separator to separate Blender’s own command-line arguments from those intended for the script.
The following is an example of how to run blender in the background to render an image of a Gaussian hill terrain,
\begin{lstlisting}[frame=single, escapechar={|}]
blender --python generate_terrain.py --background -- --save-dir folder --modules Gaussian Ground Render
\end{lstlisting}
All the presented modules work when running the scripts from Blender, in addition to a few \emph{Blender} specific modules.
These can be used to render images, but also to e.g. embed meshes in a terrain or generate segmentation masks.
We first describe the use for visualisation.

The module \texttt{Ground} creates grids from the terrains.
By default, these are colored according to height with a grayscale colormap.
The \texttt{ColorMap} module can be used to change the colormap, using syntax like \texttt{ColorMap:coolwarm}.
This supports colormaps from both \emph{ColorCet} \cite{kovesi2015good} and \emph{matplotlib} \cite{hunter2007matplotlib}.
The \texttt{ImageTexture} module can instead be used to apply a texture to the terrain, using the syntax \texttt{ImageTexture:path-to-image}.
A texture can also be supplied through the pipe by certain plot modules (e.g., \texttt{PlotObstacles}), if the setting \texttt{exportmode:True} is used.
This produces versions of the plots without axes, allowing them to be properly mapped onto the terrain.

The \texttt{Camera} module sets up a camera, at an angle relative the z-axis with the syntax \texttt{Camera:angle}, and \texttt{Render} renders an image, possibly taking a filename or folder as input with the syntax \texttt{Render:file.png} or \texttt{Render:folder}.

The camera module can also take a string input: \texttt{'top'}, which creates an \emph{orthographic}, top-down-view camera with the same dimension and resolution as the terrain.
This can be used together with the \texttt{Depth} module to generate a new terrain element, e.g. with arbitrary objects embedded into the terrain.
\texttt{Depth} renders and saves a depth image, but also use the (negative) image to create a terrain which is added to the \emph{temporary} list.

\begin{figure}[ht]
  \centering
  \includegraphics[width=0.85\columnwidth]{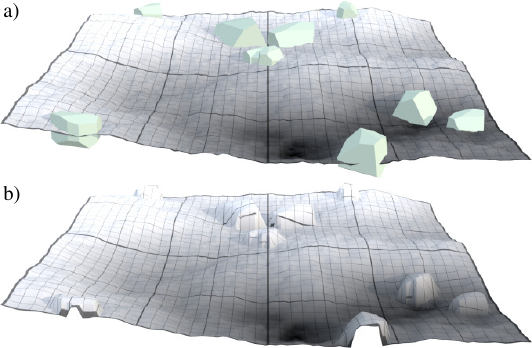}
  \caption{\label{fig:blender} a) Render of terrain and mesh-rocks created using \texttt{AddMeshObjects}, b) the same, with the rocks embedded in the terrain using \texttt{Camera:top} and \texttt{Depth}.}
\end{figure}

The \texttt{AddMeshObjects} module can add mesh-objects, given a list of obstacles, which can be embedded in a terrain using \texttt{Camera:top} and \texttt{Depth}, as seen in Fig.~\ref{fig:blender}.
The terrain is the same as in Fig.~\ref{fig:octaves_combined} and the obstacles the same as in Fig.~\ref{fig:obstacles}.
\texttt{AddMeshObjects} ignores the \texttt{height} parameter and instead places the rocks in relation to the surface height.

\begin{figure}[ht]
  \centering
  \includegraphics[width=0.85\columnwidth]{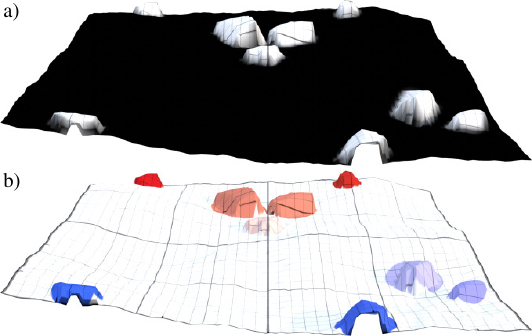}
  \caption{\label{fig:blender2} a) Render of terrain with embedded mesh-rocks with segmentation texture, b) the same, with a obstacle plot texture.}
\end{figure}

The \texttt{RenderSegmentation} module renders an image with terrain and rocks segmented with a black-and-white mask, which e.g. could be of use in applications of object identification.
Fig.~\ref{fig:blender2} shows the same terrain with embedded mesh-rocks as in Fig.~\ref{fig:blender}a, with a segmentation image and a obstacle plot used as texture.

\subsection{Control Flow}
\label{subsec:multiple_terrains}
Given a combination of modules that generates desirable terrains, multiple instances can easily be generated using multiple pipes, simply by adding \texttt{Loop} at the beginning of the list of modules:
\begin{lstlisting}[frame=single, escapechar={|}]
|\gray{python}| |\gray{generate\_terrain.py}| |\gray{--save-dir}| |\gray{folder}| |\gray{--modules}| Loop:10 ...
\end{lstlisting}
Here, the \texttt{Loop} module takes an integer input with the syntax \texttt{Loop:N}, where \texttt{N} specifies how many times the entire pipeline should be repeated.

\begin{figure}[ht]
  \centering
  \includegraphics[width=0.85\columnwidth]{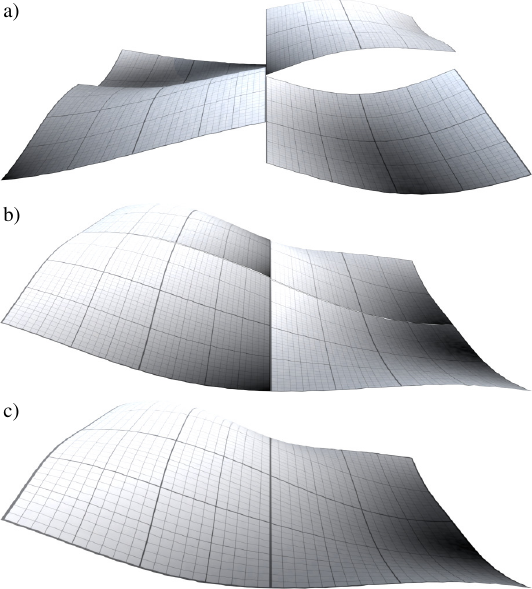}
  \caption{\label{fig:loop} a) Render of 4 terrains with adjacent \texttt{extent} that do not fit together, b) using \texttt{Seed} to make the terrains seamless, c) using \texttt{Stack} to merge the 4 terrains into one.}
\end{figure}

However, \texttt{Loop} can also be used to generate a grid of terrains that can be assembled into a larger composite result, with the syntax \texttt{Loop:AxB} where \texttt{A} and \texttt{B} are two integers.
This can be useful for creating an extended landscape without being limited by memory or computational constraints.
For example:
\begin{lstlisting}[frame=single, escapechar={|}]
|\gray{python}| |\gray{generate\_terrain.py}| |\gray{--save-dir}| |\gray{folder}| |\gray{--modules}| Loop:2x2 Basic WeightedSum Save Plot
\end{lstlisting}
This divides the \texttt{extent} into \texttt{A} pieces along the x-axis and \texttt{B} pieces along the y-axis, with the subsequent modules applied to each sub-extent in turn.
If the terrains are generated from random noise, they will not automatically fit together seamlessly, as illustrated by the terrains generated by this command and shown in Fig.~\ref{fig:loop}a.

If the aim is for the  terrains to fit together seamlessly we must set the random seed such that this is the same each time a module is called.
This can be done by inserting the \texttt{Seed} module, with the syntax \texttt{Seed:X} where \texttt{X} is some integer,
\begin{lstlisting}[frame=single, escapechar={|}]
|\gray{python}| |\gray{generate\_terrain.py}| |\gray{--save-dir}| |\gray{folder}| |\gray{--modules}| Loop:2x2 Seed:2 Basic WeightedSum Save Plot
\end{lstlisting}
as can be seen in Fig.~\ref{fig:loop}b.
As this command generates the same terrain every time, there is also the syntax \texttt{Seed:persistent\_random}, which generates a random seed the first time the module is called and reuses it thereafter.

The command above generates 4 terrains, even if they are able to fit together.
If we want to actually merge them into a single terrain, this can be done using the \texttt{Stack} module.
However, we do not want this operation to occur \emph{within the loop}, but rather \emph{after the loop}, once all 4 terrains have been generated.
To achieve this, we use it in conjunction with \texttt{EndLoop}, which is used to mark the end of a loop:
\begin{lstlisting}[frame=single, escapechar={|}]
|\gray{python}| |\gray{generate\_terrain.py}| |\gray{--save-dir}| |\gray{folder}| |\gray{--modules}| Loop:2x2 Seed:2 Basic WeightedSum EndLoop Stack Save Plot
\end{lstlisting}
This generates a single, combined terrain as seen in Fig.~\ref{fig:loop}c.

\begin{figure}[ht]
  \centering
  \includegraphics[width=0.85\columnwidth]{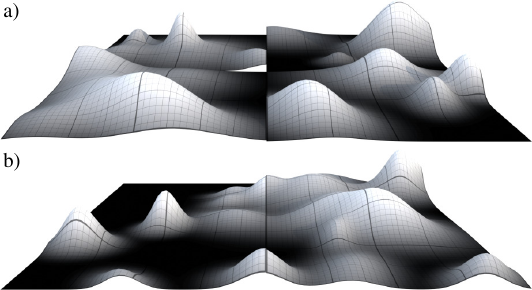}
  \caption{\label{fig:loop2} a) Render of 4 terrains with Gaussian hills from separate obstacle lists that do not fit together, b) using a common obstacle list to make the terrains seamless.}
\end{figure}

Modules that generate terrains from functions can also be used while looping over sub-extents, and can be made seamless as well.
However, care must be taken: if lists of \emph{obstacles} are generated separately for each sub-extent, edge effects may occur, as illustrated in Fig.~\ref{fig:loop2}a.
A simple solution is to generate or load all obstacles before the loop, as in Fig.~\ref{fig:loop2}b, so that the same set of obstacles is used for all sub-extents.
This allows the same obstacle to contribute to multiple sub-extents, resolving the edge effect issue.
However, it may also introduce many unnecessary obstacles, especially when generating large grids, since all obstacles are passed to every sub-extent regardless of distance.
In such cases, the \texttt{RemoveDistantObstacles:D} module can be used to discard obstacles beyond a specified distance \texttt{D}.

We can also loop over a list of values, specified either explicitly or using a deterministic sequence, in the following ways:
\begin{lstlisting}[frame=single, escapechar={|}]
Loop:parameter=[1,2,4]
Loop:parameter=arange[1,11]
Loop:parameter=linspace[0,0.5,6]
Loop:parameter=logspace[-1,1,10]
\end{lstlisting}
This is similar to \texttt{Loop:N}, except that the parameter takes on explicitly specified values, which can influence subsequent modules.
Multiple loops can also be stacked sequentially.
We will demonstrate this in section~\ref{sec:results}, as shown in Listing~\ref{lst:double_loop}.

\subsection{Module}
\label{subsec:module}
Here we go into details of what a module is and how it operates within the framework of \texttt{generate\_terrain.py}.
Modules are designed to operate in two different modes.
The simple description is that a module is an object with a callable method that takes input via keyword arguments and returns a dictionary, which is combined with the \emph{pipe} dictionary and passed as input to the next module.
This description is sufficient knowledge for developing new modules within the framework: all that is needed is to implement the callable method.
The reason we need to go beyond this simple explanation is the setup with multiple pipes using \texttt{Loop}.

In \texttt{generate\_terrain.py}, module execution follows a \emph{depth-first} approach, where the full list of modules is executed for one iteration of a \texttt{Loop}, before proceeding to the next.
To achieve this, we utilize \emph{recursion}.
In this context, a module is an \emph{iterable} object, which executes and \emph{yields} a dictionary when iterated over.
As for the simple description, this dictionary is combined with the \emph{pipe} and used as input to the next module, which is called recursively.
By default, a module yields only once and runs the same callable method described earlier, making the two descriptions identical.

\texttt{Loop} is simply a module with a custom \texttt{loop\_generator} method, which yields multiple times instead of the default $1$.
\texttt{EndLoop}, in contrast, suppresses iteration until the final loop iteration, allowing intermediate values to be collected and merged before passing control to the next module.

Formally this is implemented in python as a \emph{generator}.
Modules are initialised before any call to allow them to retain memory between invocations.
When control is passed from one module to the next, a \texttt{start} method is run with the purpose of collecting the keyword input and initialising the generator function.
The setup of \texttt{Module} can be seen in Listing~\ref{lst:module}, and the recursive setup of \texttt{generate\_terrain.py} can be seen in Listing~\ref{lst:recursive}.
\begin{lstlisting}[
    frame=single,
    caption={Slightly compacted definition of a \texttt{Module}.},
    label={lst:module},
  ]
class Module():
    def __init__(self, save_dir=None, logger=None):
        # Setup save-dir and logger
        ...

    def __call__(self, **kwargs):
        raise NotImplementedError

    def start(self, **kwargs):
        # Store kwargs
        self.kwargs = kwargs
        # Initialize loop generator
        self.loop_generator_instance = self.loop_generator()

    def loop_generator(self):
        # yield only once, using output from __call_
        yield self.__call__(**self.kwargs)

    def __iter__(self):
        return self.loop_generator_instance
\end{lstlisting}

\begin{lstlisting}[label=lst:recursive,caption=Slightly compacted definition of the recursive setup of \texttt{generate\_terrain.py}.,
    frame=single,
]
def recursive_module_call(
        list_of_modules_kwargs_tuples, index=0, pipe={}):
    '''
    Call next level recursively
    '''
    # Base case: reached the end of the module list
    if index >= len(list_of_modules_kwargs_tuples):
        return None

    # Get module and kwargs
    module_obj, kwargs = list_of_modules_kwargs_tuples[index]

    module_obj.start(**kwargs, **pipe)
    for returned_data in module_obj:
        if isinstance(returned_data, dict):
            pipe = {**pipe, **returned_data}

        recursive_module_call(
            list_of_modules_kwargs_tuples, index+1, pipe)
\end{lstlisting}

\section{Results}
\label{sec:results}
Here we present results focusing on terrain and rock parametrization, demonstrating how the modular system can be used to generate, modify, and analyze terrain features.

\subsection{Parameterized terrains}
\label{subsec:parameterized_terrains}
We aim to enable the generation of terrains based on specific parameters that describe the traversability of the terrain, e.g. in terms of measures like \emph{slope} and \emph{roughness}.
Since a given set of terrain elements can be combined into arbitrarily easy or difficult terrain depending on how they are weighted, the weighting becomes a natural target for parameterization.
Below, we demonstrate how to control the slope and roughness (approximately) of a resultant terrain by adjusting the weights when combining multiple \texttt{Octaves} using a \texttt{WeightedSum}.

The idea is as follows: Given that we want the resultant terrain to achieve a specific value for a certain measure, we 1) calculate the measure for each terrain element, 2) find an exact or approximate expression (a proxy measure) for how these combine for the resultant terrain, and 3) adjust the weights so that the proxy measure reaches the desired value.
By scaling the weights in proportion to the proxy measure values, we ensure minimal changes to components that have little influence on the final result.
For example, surface roughness will mostly depend on terrains with small-scale variations, whereas slope will primarily depend on terrains with large-scale variations, so modifying one should not significantly affect the other.

There is no single value for slope from which one can estimate the resultant slope, as terrains with a given inclination can interfere constructively or destructively.
However, by using the \emph{gradient}, we can account for slope in both the x- and y-directions, and the resultant gradient can be computed exactly as
\begin{equation}
  \label{}
  \nabla z = \sum_{i=1} w_i \nabla z_i
\end{equation}
where $w_i$ are the weights and $\nabla z_i$ the gradients of the terrain elements.
We define the \emph{mean gradient} $\bar{\nabla} z$ as the average of the gradient vectors over all points in the terrain,
\begin{equation}
  \bar{\nabla} z = \frac{1}{N} \sum_{x,y} \nabla z_{x,y}
\end{equation}
where $N$ is the number of points, and we define the \emph{slope} as
\begin{equation}
s = \tan^{-1}(\vert \bar{\nabla} z \vert).
\end{equation}

We define the (relative surface area) \emph{roughness} as the ratio between the actual surface area of the terrain and the area of a smoothed version of the same terrain:
\begin{equation}
r = \frac{A_\text{orig}}{A_\text{smooth}}
\end{equation}
A perfectly flat terrain would have a roughness of 1, while a highly irregular or jagged surface would yield a higher value.

\begin{figure}[ht]
  \centering
  \includegraphics[width=0.95\columnwidth]{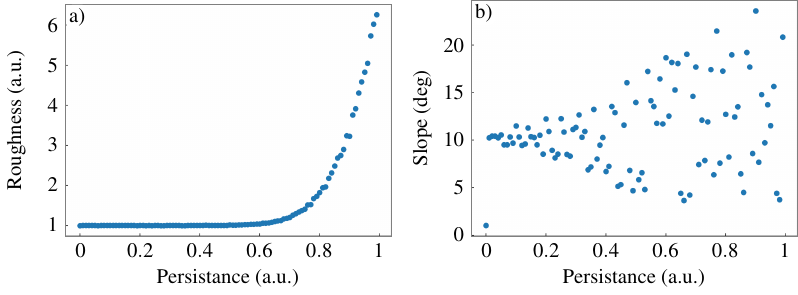}
  \caption{\label{fig:roughness_and_slope} a) Roughness and b) Slope of a \texttt{WeightedSum} of \texttt{Octaves} for varying \texttt{persistence}.}
\end{figure}

\begin{figure}[ht]
  \centering
  \includegraphics[width=0.95\columnwidth]{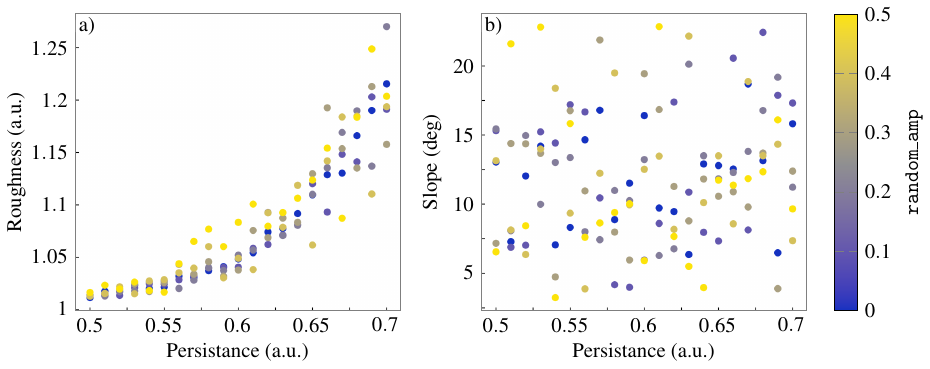}
  \caption{\label{fig:roughness_and_slope_2} a) Roughness and b) Slope of a \texttt{WeightedSum} of \texttt{Octaves} for varying \texttt{persistence} for 6 different levels of \texttt{random\_amp}. }
\end{figure}

The module \texttt{Slope} calculates the slope $s_i$ and the mean gradient $\bar{\nabla} z_i$ for all terrain elements.
The module \texttt{Roughness:sigma\_meter} calculates roughness $r_i$ using a Gaussian smoothing with standard deviation given by \texttt{sigma\_meter}, which by default is $5$~m.
We will use the measures above to parameterize terrains, but first we use them to examine the parameters in the \texttt{Octaves} module.
As described earlier, \texttt{Octaves} generates a list of terrain elements where the spatial scale is halved for each successive terrain, along with a corresponding list of \texttt{weights}.
These are derived from a starting amplitude (\texttt{amplitude\_start} = 10), which is scaled down by a factor (\texttt{persistence} = 0.60) for each subsequent terrain.
With this setup, the slope and roughness measures are partly connected.
A high \texttt{persistence} value results in higher roughness and can also lead to higher slope values, though slope tends to vary more due to constructive or destructive interference.
This relationship is illustrated in Fig.~\ref{fig:roughness_and_slope}, where we generate 100 terrains while sweeping persistence values from 0.01 to 1.0,
To introduce greater variation, there is a parameter \texttt{random\_amp} in the \texttt{Octaves} module, where each weight is multiplied by a random factor drawn from $\sim \mathcal{N}(1, \texttt{random\_amp})$, as well as assigned a random sign.
The relationship between persistence, roughness, and slope for different levels of \texttt{random\_amp} and with \texttt{random\_sign} is shown in Fig.~\ref{fig:roughness_and_slope_2}, over a more narrow and practically useful range.
Even though the variation increases, we observe a clear relationship between persistence and roughness, whereas the relationship between persistence and slope appears random.

We now move on to parametrizing the terrains.
We construct the module \texttt{SetSlope} to alter the weights for \texttt{WeightedSum} given a \texttt{target\_slope\_deg} input, such that a terrain with mean slope of $10^{\circ}$ can be constructed as
\begin{lstlisting}[frame=single, escapechar={|}]
|\gray{--modules}| Octaves Slope SetSlope:10 WeightedSum
\end{lstlisting}

For roughness, we cannot derive an exact expression for the resultant roughness as a weighted sum of the individual roughness measures, but we may look for an approximation.
We consider an expression where a smooth terrain ($r=1$) gives no contribution, and introduce $\alpha$ and $\beta$ as scaling parameters,
\begin{equation}
  \label{}
r_{\text{proxy}} = 1 + \left[ \sum_i (r_i^{\alpha}-1) w_i^{\beta} \right]^{1/\alpha},
\end{equation}
with the aim of finding values of $\alpha$ and $\beta$ that yield a reasonable approximation of the actual resultant roughness.
We generate terrains using \texttt{Octaves} with varying \texttt{weights} and different numbers of components, and find that values close to $\alpha=1$ and $\beta=2$ provide a reasonably good fit.

\begin{figure}[ht]
  \centering
  \includegraphics[width=0.55\columnwidth]{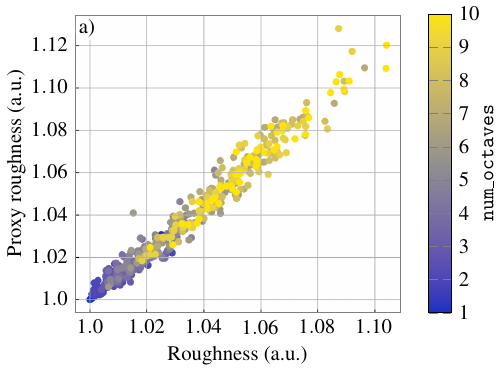}
  \caption{\label{fig:roughness_v_proxy} Scatter plot of roughness v \emph{proxy roughness} for terrains created with \texttt{Octaves} with \texttt{num\_octaves} between 1 and 10, combined using \texttt{WeightedSum} with varying weights.}
\end{figure}

We compare the roughness and the \emph{proxy roughness} $r_{\text{proxy}}$ for a total of 500 terrains, each composed using the \texttt{Octaves} module with a varying number of octaves and weights perturbed by a random factor (\texttt{random\_amp = 0.5}).
The results are shown in Fig.~\ref{fig:roughness_v_proxy}.
Similar to \texttt{SetSlope}, we construct the module \texttt{SetRoughness} to alter the weights for \texttt{WeightedSum} to achieve a proxy roughness, given a \texttt{target\_roughness} input.

\begin{figure}[ht]
  \centering
  \includegraphics[width=0.95\columnwidth]{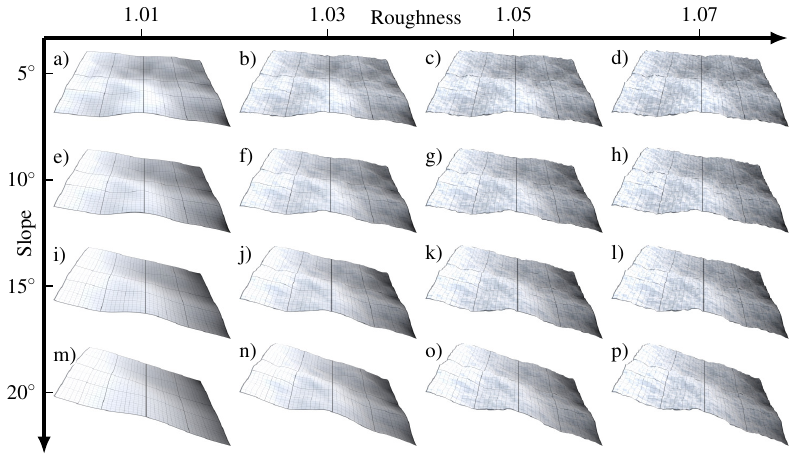}
  \caption{\label{fig:roughness_and_slope_render_grid} Rendered images of examples where slope and roughness are varied, for the same set of terrain elements.}
\end{figure}

We test \texttt{SetSlope} and \texttt{SetRoughness} in combination by loading the same saved terrain elements and combining them for different slope and roughness, as seen in Fig.~\ref{fig:roughness_and_slope_render_grid}.

We note that the target values for \texttt{SetSlope} and \texttt{SetRoughness} can be sampled from distributions, allowing the generation of ensembles of terrains with desired characteristics.
The parameterization of slope and roughness can be seen as an example, and it would be possible to develop similar approaches for other measures as well.
In the Swedish forestry industry, for example, there are established classification systems for slope and surface roughness used to evaluate areas prior to different types of logging operations \cite{berg1982terrangtypschema}.
In the following section, we consider how the parameters of the \texttt{Rocks} module can be parameterized, e.g. to achieve different surface roughness according to \textcite{berg1982terrangtypschema}.

\subsection{Parameterized rocks}
We can generate rocks of varying sizes using the \texttt{Rocks} module.
The module is based on simplex noise of different length scales, where the peaks are truncated and rescaled to resemble rocks.
It takes a list of \texttt{rock\_size} as input (by default \texttt{[0.5, 1, 2, 4]}), representing the length scale.
The number of rocks can be adjusted using the \texttt{fraction} parameter, and their heights can be controlled via \texttt{rock\_heights}.
In this analysis, however, we retain the default setting, where \texttt{rock\_heights} is set to half the corresponding length scale.

Given a generated terrain-basic $z_{xy}$ with theoretical max-value $z_{\text{max}}$ and \texttt{fraction}~$\in [0, 1]$, we select all points where $z_{xy} > \texttt{fraction} \times z_{\text{max}}$.
We rescale these such that the interval $[\texttt{fraction} \times h_{\text{max}}, h_{\text{max}}]$ is mapped to $[0, \texttt{rock\_heights}]$, and set all other points to 0.

In the previous section on terrain parameterization, we focused on how terrains of different scales were weighted to achieve specific properties of the combined terrain.
Here, rocks are directly assigned reasonable heights, and we parameterize them using \texttt{rock\_size} and \texttt{fraction}.
We analyse the resulting terrains using the \texttt{FindRocks} and \texttt{SurfaceStructure} modules.
\texttt{SurfaceStructure} classifies the list of obstacles into four height categories, $\text{h}_{20}, \text{h}_{40}, \text{h}_{60}, \text{h}_{80}$, representing the number of rocks in the height intervals $[10, 30]$~cm, $[30, 50]$~cm, $[50, 70]$~cm, and $[70, \infty]$~cm, respectively.
It also computes a single surface roughness measure, $Y \in [1, 5]$, indicating surface roughness from “very smooth” to “very rough”, according to \cite{berg1982terrangtypschema}.

\begin{figure}[ht]
  \centering
  \includegraphics[width=0.55\columnwidth]{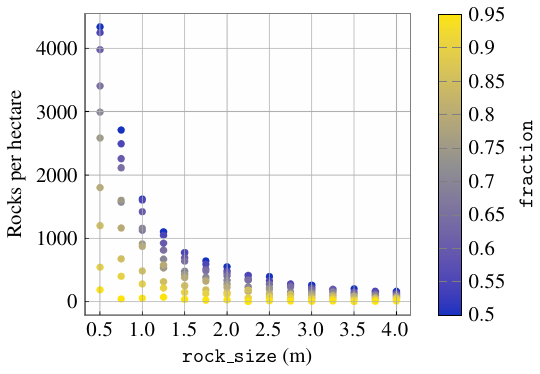}
  \caption{\label{fig:num_rocks_1} Number of rocks per hectare, given different values of the \texttt{Rocks} parameters \texttt{rock\_size} and \texttt{fraction}.}
\end{figure}

\begin{figure}[ht]
  \centering
  \includegraphics[width=0.95\columnwidth]{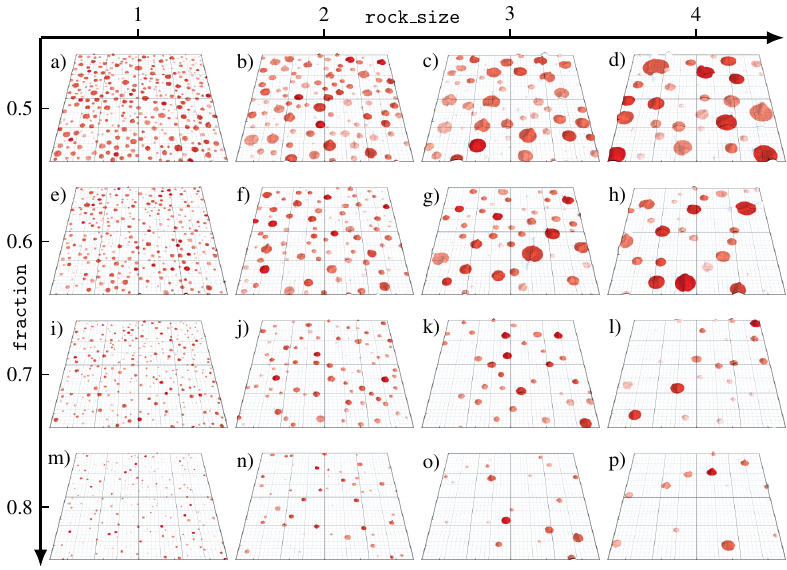}
  \caption{\label{fig:num_rocks_grid} Rendered images of examples where \texttt{rock\_size} and \texttt{fraction} are varied.}
\end{figure}

We analyse the input parameters by generating rock terrains using different values of \texttt{rock\_size} and \texttt{fraction}, and count the resultant number of rocks.
The result can be seen in Fig.~\ref{fig:num_rocks_1}, with Fig.~\ref{fig:num_rocks_grid} showing rendered terrain examples with texture from \texttt{PlotObstacles} in the form of red disks at each identified rock.
There is a cut-off such that rocks below $10$~cm of height are not counted.
As we increase \texttt{rock\_size}, the rocks become fewer but larger.
An increase by a factor of 2 might be expected to result in a factor of 4 fewer rocks, but this is partially mitigated by two factors: (i) fewer rocks falling below the $10$~cm height threshold, and (ii) a boundary effect, whereby larger rocks are more likely to partially intersect the terrain.
We observe that both the number and size of rocks decrease as \texttt{fraction} increases, since a smaller portion of the noise peaks is used.
Also, in the above examples we have employed a higher terrain resolution of $0.1 \times 0.1$~m to resolve smaller rocks.

\begin{figure}[ht]
  \centering
  \includegraphics[width=0.95\columnwidth]{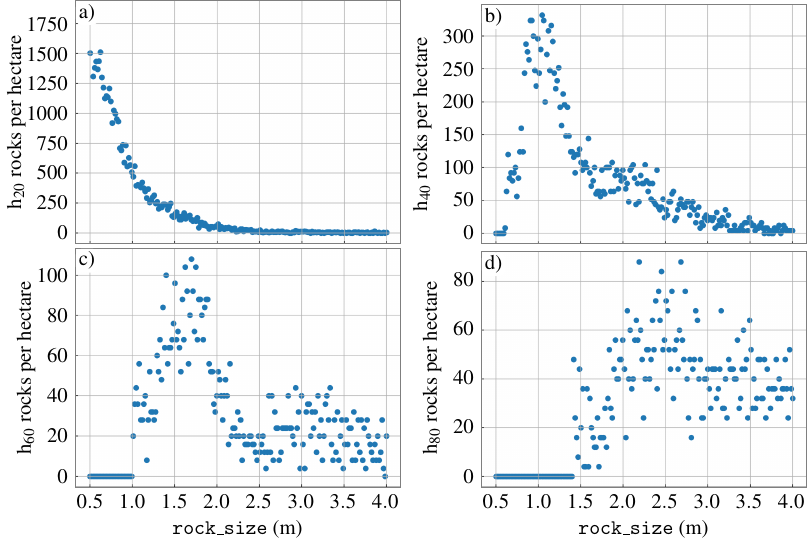}
  \caption{\label{fig:num_rocks_2} Number of rocks per hectare in the 4 height intervals $\text{h}_{20}$, $\text{h}_{40}$, $\text{h}_{60}$, $\text{h}_{80}$, as a function of \texttt{rock\_size}.}
\end{figure}

Different values of \texttt{rock\_size} contribute differently to the four height intervals, as shown in Fig.~\ref{fig:num_rocks_2}.
It is thus possible to construct a desired distribution of rocks, in terms of both size and quantity, by selecting the appropriate list of \texttt{rock\_size}.
To achieve an even greater number of rocks of a given size, it is of course possible to combine multiple instances of the same configuration.

\begin{figure}[ht]
  \centering
  \includegraphics[width=0.55\columnwidth]{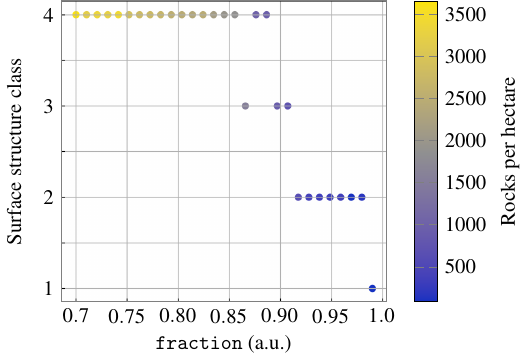}
  \caption{\label{fig:num_rocks_3} Scatter plot of resultant roughness class for different values of \texttt{fraction}, for default \texttt{rock\_size} using \texttt{Rocks}.}
\end{figure}

We investigate the resulting surface roughness class for the default values of \texttt{rock\_size}, $[0.5, 1, 2, 4]$, as we vary the \texttt{fraction} parameter.
The results are shown in Fig.~\ref{fig:num_rocks_3}, with color coding indicating the number of rocks.

\begin{figure}[ht]
  \centering
  \includegraphics[width=0.95\columnwidth]{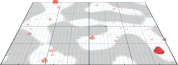}
  \caption{\label{fig:num_rocks_spatial} Default terrain of \texttt{Rocks}, multiplied by a \texttt{Basic:10} mask to create a spatial distribution where there are no rocks in the gray regions.}
\end{figure}

The spatial distribution of rocks can also be influenced by multiplying the generated rock terrain by another array, i.e., by applying a secondary terrain as a multiplicative mask.
An example is shown in Fig.~\ref{fig:num_rocks_spatial}, where the rocks are modulated by a \texttt{Basic:10} terrain, scaled by a factor of $10$ and clipped to the range $[0, 1]$, resulting in plateau regions with values of $0$ and $1$.
This could, of course, serve as an alternative way of parameterizing the rocks, for example by vertically shifting the mask before clipping, thereby controlling the resulting proportion of values at $0$ and $1$.

\begin{figure}[ht]
  \centering
  \includegraphics[width=0.55\columnwidth]{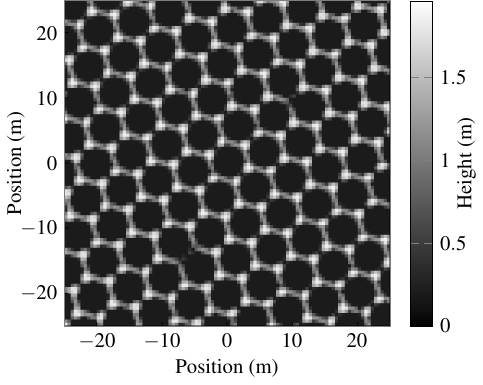}
  \caption{\label{fig:num_rocks_hexagon} Plot of 1000 \texttt{Rocks:4} terrains combined via \texttt{Combine:Max}, to show restrictions in rock positions, unless \texttt{random\_shift} is enabled.}
\end{figure}

On a final note: Due to the properties of simplex noise, there are inherent constraints on where maxima and minima can occur, meaning that rocks cannot appear completely uniformly across the terrain.
This effect can be observed in Fig.~\ref{fig:num_rocks_hexagon}, where we have generated 1000 \texttt{Rocks:4} terrains and combined them using a \texttt{Combine:Max} operation.
However, this limitation can be easily addressed by introducing an appropriate random shift, implemented in \texttt{Rocks} using the Boolean parameter \texttt{random\_shift}.
By linking this shift to the random seed used for the simplex noise, we can ensure that the same shift is applied for a given seed, enabling reproducibility and consistent \texttt{Stack} operations.
The same analysis can be applied to \texttt{Holes}, which are equivalent to \texttt{Rocks} but with negative and halved amplitude.

\section{Discussion}
\label{sec:discussion}
The results illustrate how terrain characteristics can be parametrized using the modular system, and how the same approach can be extended to support additional metrics.
The library is designed to allow users to define new modules or parametrization methods to meet specific needs, supporting both rapid prototyping and long-term development.

Compared to other tools, our focus is not on visual realism or artist-driven workflows.
Instead, we target simulation-driven research and applications where control, reproducibility, and parameterization are central, such as generating synthetic terrains for machine learning or ground truth for segmentation tasks.

One of the key strengths of the system is the compact and flexible architecture: a small set of modules can be combined in numerous ways to produce complex results.
This is possible due to the consistent interface and the general-purpose design of the modules.
For example, a generated terrain can be reused as a mask, sampling distribution, lookup map, or scaling factor.
Likewise, plots can be used for both analysis and texturing, while obstacle lists, in addition to placing obstacles, can be used to generate function-based terrains or serve as segmentation output.
Distributions serve multiple purposes, from sampling obstacle lists to generating parameter inputs for other modules.
As a result, a new module or feature will often find multiple uses, enabling a system whose capabilities scale favorably with each addition.

There are, however, limitations.
The system is based on 2D heightmaps and cannot represent overhanging structures.
While Blender can be used to include arbitrary geometry, the integration is currently focused on embedding such geometry into terrains rather than handling them as independent, fully featured objects.

The command-line and settings-file interface supports reproducibility, scripting, and integration with automated pipelines, while making it easy to share and version-control complete configurations.
Although this approach may present a steeper learning curve for some users, the modular structure lends itself well to a visual programming interface.
Such an interface could form the basis of a web-based front-end, lowering the entry threshold while preserving the system’s flexibility.

We have focused on demonstrating a flexible system with the desired capabilities using a minimal set of modules.
However, the design readily supports extension, and many valuable additions could be implemented with relatively little effort.

\section{Conclusions}
\label{sec:conclusion}
We conclude that our library for parameterized procedural terrain generation offers the flexibility and control needed to support simulation-driven development of intelligent machines in outdoor environments.
We demonstrate that our modular pipeline can generate complex landscapes with explicit control over key characteristics.

Our main contribution is the software library itself, released under the MIT license at \url{https://github.com/erikwallin86/artificial_terrains}, and our key findings include that
\begin{itemize}
  \item the proxy-based weighting strategy for slope and roughness enables parameterization across a wide range of terrain configurations;
  \item the configurable rock-generation parameters allow the user to achieve a desired distribution of the surface roughness class; and
  \item the integration with Blender supports high-quality rendering, seamless embedding of mesh objects, and generation of segmentation masks for downstream perception tasks.
\end{itemize}

The modular interface of the library makes it straightforward to extend and build upon existing functionality.
Future work includes new terrain analysis metrics, support for additional file formats, and metadata for terrain attributes.
It would also be valuable to develop modules for other typical terrain elements beyond rocks and holes, as well as to demonstrate how physics-driven processes or AI-based generators can be integrated into the system.
Moreover, additional mathematical, modifier, and composition modules could be implemented to expand the range of supported operations.
Future work could explore deeper Blender integration and richer file format support to extend beyond heightfields.
Finally, a visual programming interface, possibly web-hosted, would lower the barrier to entry and enable users to assemble and configure terrain pipelines in a drag-and-drop environment, encouraging rapid prototyping and experimentation.

\section{Acknowledgements}
This work was supported by the Åforsk Foundation (grant no.\ 23-273) and Mistra Digital Forest, a research program funded by the Swedish Foundation for Strategic Environmental Research (Mistra) and the Swedish forestry industry.

The author would like to thank Martin Servin, Arvid Fälldin, and Mikael Lundbäck for their input on the design and implementation of the procedural terrain modules and for their thoughtful feedback.

\printbibliography

\onecolumn
\appendix

\section{Appendix}

\subsection{Commands for figures}
\label{subsec:commands_for_figures}
Even if the main sections include many of the commands needed to generate the terrains in the figures, it is still not exactly the commands run to e.g. \emph{render} the images.
The following section contains the full commands for generating all terrains and figures which are presented.
\begin{lstlisting}[frame=single, escapechar={|},
caption={Full commands for generating \texttt{Basic} Figs.~\ref{fig:basic}-\ref{fig:basic_combined}},
]
|\gray{python}| |\gray{generate\_terrain.py}| |\gray{--save-dir}| |\gray{basic}| |\gray{--modules}| Basic Plot Save
|\gray{blender}| |\gray{--python}| |\gray{generate\_terrain.py}| |\gray{--}| |\gray{--save-dir}| |\gray{basic}| |\gray{--modules}| Load:basic/Save WeightedSum Ground Camera:75 Render
|\gray{blender}| |\gray{--python}| |\gray{generate\_terrain.py}| |\gray{--}| |\gray{--save-dir}| |\gray{basic}| |\gray{--modules}| Load:basic/Save WeightedSum:[10,3,0.3] Ground Camera:75 Render:b.png
\end{lstlisting}

\begin{lstlisting}[frame=single, escapechar={|},
caption={Full commands for generating \texttt{Octaves} Figs.~\ref{fig:octaves}-\ref{fig:octaves_combined}},
]
|\gray{python}| |\gray{generate\_terrain.py}| |\gray{--save-dir}| |\gray{octaves}| |\gray{--modules}| Octaves Plot Save
|\gray{blender}| |\gray{--python}| |\gray{generate\_terrain.py}| |\gray{--}| |\gray{--save-dir}| |\gray{octaves}| |\gray{--modules}| Load:octaves/Save Random:weights WeightedSum Save:Save2 Ground Camera:75 Holdout Render
|\gray{blender}| |\gray{--python}| |\gray{generate\_terrain.py}| |\gray{--}| |\gray{--save-dir}| |\gray{octaves}| |\gray{--modules}| Load:octaves/Save Random:weights WeightedSum Ground Camera:75 Holdout Render:b.png
\end{lstlisting}

\begin{lstlisting}[frame=single, escapechar={|},
caption={Full commands for generating \texttt{Rocks} Figs.~\ref{fig:rocks}-\ref{fig:rocks_combined}},
]
|\gray{python}| |\gray{generate\_terrain.py}| |\gray{--save-dir}| |\gray{rocks}| |\gray{--modules}| Rocks Plot Save
|\gray{blender}| |\gray{--python}| |\gray{generate\_terrain.py}| |\gray{--}| |\gray{--save-dir}| |\gray{rocks}| |\gray{--modules}| Load:rocks/Save Combine Ground Camera:75 Render
|\gray{blender}| |\gray{--python}| |\gray{generate\_terrain.py}| |\gray{--}| |\gray{--save-dir}| |\gray{rocks}| |\gray{--modules}| Load:rocks/Save Combine Load:octaves/Save2 Combine Ground Camera:75 Render:with_terrain.png
\end{lstlisting}

\begin{lstlisting}[frame=single, escapechar={|},
    caption={Full commands for generating \texttt{Holes} Fig.~\ref{fig:holes_combined}},
]
|\gray{python}| |\gray{generate\_terrain.py}| |\gray{--save-dir}| |\gray{holes}| |\gray{--modules}| Holes Plot Save
|\gray{blender}| |\gray{--python}| |\gray{generate\_terrain.py}| |\gray{--}| |\gray{--save-dir}| |\gray{holes}| |\gray{--modules}| Load:holes/Save Combine Ground Camera:75 Render
|\gray{blender}| |\gray{--python}| |\gray{generate\_terrain.py}| |\gray{--}| |\gray{--save-dir}| |\gray{holes}| |\gray{--modules}| Load:holes/Save Combine Load:octaves/Save2 Combine Ground Camera:75 Render:with_terrain.png
\end{lstlisting}

\begin{lstlisting}[frame=single, escapechar={|},
    caption={Full command for generating Fig.~\ref{fig:generating_function}},
]
|\gray{python}| |\gray{generate\_terrain.py}| |\gray{--save-dir}| |\gray{gen\_func}| |\gray{--modules}| Gaussian Step Donut Plane Sphere Cube SmoothStep Sine Plot Save
\end{lstlisting}

\begin{lstlisting}[frame=single, escapechar={|},
    caption={Full commands for generating Fig.~\ref{fig:generating_function_2}},
]
|\gray{blender}| |\gray{--python}| |\gray{generate\_terrain.py}| |\gray{--}| |\gray{--save-dir}| |\gray{gen\_fun\_2}| |\gray{--modules}| Random:10 Gaussian Combine Ground Camera:75 Render
|\gray{blender}| |\gray{--python}| |\gray{generate\_terrain.py}| |\gray{--}| |\gray{--save-dir}| |\gray{gen\_fun\_2}| |\gray{--modules}| SetDistribution:'height=uniform(-5,5)' Random:10 Gaussian Combine Ground Camera:75 Render:b.png
\end{lstlisting}

\begin{lstlisting}[frame=single, escapechar={|},
    caption={Full commands for generating Figs.~\ref{fig:distributions}-\ref{fig:obstacles}},
]
|\gray{blender}| |\gray{--python}| |\gray{generate\_terrain.py}| |\gray{--}| |\gray{--save-dir}| |\gray{2Dsampling}| |\gray{--modules}| Donut:"dict(width=40)" AsProbability Random:10 Sphere Combine Ground Camera:75 Render
|\gray{blender}| |\gray{--python}| |\gray{generate\_terrain.py}| |\gray{--}| |\gray{--save-dir}| |\gray{2D\_lookup}| |\gray{--modules}| Plane:"dict(pitch_deg=10)" AsLookupFor:height Random:10 SaveObstacles Gaussian Combine Ground Camera:75 Render:lookup.png
|\gray{python}| |\gray{generate\_terrain.py}| |\gray{--save-dir}| |\gray{plot\_obstacles}| |\gray{--modules}| LoadObstacles:2D_lookup/SaveObstacles/obstacles.npz PlotObstacles
\end{lstlisting}

\begin{lstlisting}[frame=single, escapechar={|},
    caption={Full command for generating Fig.~\ref{fig:function}},
]
|\gray{blender}| |\gray{--python}| |\gray{generate\_terrain.py}| |\gray{--}| |\gray{--save-dir}| |\gray{function}| |\gray{--modules}| Function:'5*(x/np.max(x))**2+np.sin(y/2)' Ground Camera:65 Render
\end{lstlisting}

\begin{lstlisting}[frame=single, escapechar={|},
    caption={Full commands for generating Fig.~\ref{fig:blender}},
]
|\gray{blender}| |\gray{--python}| |\gray{generate\_terrain.py}| |\gray{--}| |\gray{--save-dir}| |\gray{blender}| |\gray{--modules}| Load:octaves/Save LoadObstacles:2D_lookup/SaveObstacles/obstacles.npz AddMeshObjects Ground Camera:75 Render Camera:top Depth Save
|\gray{blender}| |\gray{--python}| |\gray{generate\_terrain.py}| |\gray{--}| |\gray{--save-dir}| |\gray{blender}| |\gray{--modules}| Load:blender/Save Ground Camera:75 Render:combined.png
\end{lstlisting}

\begin{lstlisting}[frame=single, escapechar={|},
    caption={Full commands for generating Fig.~\ref{fig:blender2}},
]
|\gray{blender}| |\gray{--python}| |\gray{generate\_terrain.py}| |\gray{--}| |\gray{--save-dir}| |\gray{blender}| |\gray{--modules}| Load:blender/Save Ground ImageTexture:plot_obstacles/PlotObstacles/obstacles_0.png Camera:75 Render:texture1.png
|\gray{blender}| |\gray{--python}| |\gray{generate\_terrain.py}| |\gray{--}| |\gray{--save-dir}| |\gray{blender}| |\gray{--modules}| Load:octaves/Save LoadObstacles:2D_lookup/SaveObstacles/obstacles.npz AddMeshObjects Ground Camera:top RenderSegmentation:segmentation_top.png
|\gray{blender}| |\gray{--python}| |\gray{generate\_terrain.py}| |\gray{--}| |\gray{--save-dir}| |\gray{blender}| |\gray{--modules}| Load:blender/Save Ground ImageTexture:blender/RenderSegmentation/segmentation_top.png Camera:75 Render:texture2.png
\end{lstlisting}

\begin{lstlisting}[frame=single, escapechar={|},
    caption={Full commands for generating Fig.~\ref{fig:loop}. Note that when \texttt{Ground} is before \texttt{EndLoop}, a grid is created to each sub-extent in the \texttt{Loop}. When \texttt{Ground} is after \texttt{EndLoop} and \texttt{Stack}, a single grid is created for the merged terrain.},
]
|\gray{blender}| |\gray{--python}| |\gray{generate\_terrain.py}| |\gray{--}| |\gray{--save-dir}| |\gray{loop}| |\gray{--modules}| Loop:2x2 Basic WeightedSum Ground EndLoop Camera:75 Render:1.png
|\gray{blender}| |\gray{--python}| |\gray{generate\_terrain.py}| |\gray{--}| |\gray{--save-dir}| |\gray{loop}| |\gray{--modules}| Loop:2x2 Seed:2 Basic WeightedSum Ground EndLoop Camera:75 Render:2.png
|\gray{blender}| |\gray{--python}| |\gray{generate\_terrain.py}| |\gray{--}| |\gray{--save-dir}| |\gray{loop}| |\gray{--modules}| Loop:2x2 Seed:2 Basic WeightedSum EndLoop Stack Ground Camera:75 Render:3.png
\end{lstlisting}

\begin{lstlisting}[frame=single, escapechar={|},
    caption={Full commands for generating Fig.~\ref{fig:loop2}: Placing \texttt{Random} before \texttt{Loop} ensures the resultant terrains can be seamlessly combined.}
]
|\gray{blender}| |\gray{--python}| |\gray{generate\_terrain.py}| |\gray{--}| |\gray{--save-dir}| |\gray{loop}| |\gray{--modules}| Loop:2x2 Random:5 Gaussian Combine Ground EndLoop Camera:75 Render:4.png
|\gray{blender}| |\gray{--python}| |\gray{generate\_terrain.py}| |\gray{--}| |\gray{--save-dir}| |\gray{loop}| |\gray{--modules}| Random:20 Loop:2x2 Gaussian Combine Ground EndLoop Camera:75 Render:5.png
\end{lstlisting}

\begin{lstlisting}[frame=single, escapechar={|},
    caption={Full commands for generating Fig.~\ref{fig:roughness_and_slope}: \texttt{LogData} is used to save data in the loop, which is loaded and plotted in the second command.}
]
|\gray{python}| |\gray{generate\_terrain.py}| |\gray{--save-dir}| |\gray{slope\_plot}| |\gray{--modules}| Loop:persistence=linspace[0,0.99,100] Octaves:"dict(random_amp=0.0)" WeightedSum Slope Roughness LogData
|\gray{python}| |\gray{generate\_terrain.py}| |\gray{--save-dir}| |\gray{slope\_plot}| |\gray{--modules}| LoadData:slope_plot/LogData/data.npz PlotScatter
\end{lstlisting}

\begin{lstlisting}[frame=single, escapechar={|},
    caption={Full commands for generating Fig.~\ref{fig:roughness_and_slope_2}: Loop over random-amp and persistence, and plotting the data as a scatter plot, colored by the random-amp.},
    label={lst:double_loop},
]
|\gray{python}| |\gray{generate\_terrain.py}| |\gray{--save-dir}| |\gray{slope\_plot2}| --modules Loop:random_amp=linspace[0,0.5,6] Loop:persistance=linspace[0.5,0.7,21] Octaves WeightedSum Slope Roughness LogData
|\gray{python}| |\gray{generate\_terrain.py}| |\gray{--save-dir}| |\gray{slope\_plot2}| |\gray{--modules}| LoadData:slope_plot2/LogData/data.npz PlotScatter:"dict(color='random_amp',cmap='cet_bjy')"
\end{lstlisting}

\begin{lstlisting}[frame=single, escapechar={|},
    caption={Full commands for generating Fig.~\ref{fig:roughness_v_proxy}: Construct 500 terrains as a \texttt{WeightedSum} with different number of \texttt{Octaves}. Calculate both \emph{proxy roughness} and actual roughness and compare the two in a scatter plot.}
]
|\gray{python}| |\gray{generate\_terrain.py}| |\gray{--save-dir}| |\gray{proxy-roughness}| |\gray{--modules}| Loop:num_octaves=arange[1,11] Loop:50 Octaves Roughness CombineRoughness WeightedSum Roughness LogData
|\gray{python}| |\gray{generate\_terrain.py}| |\gray{--save-dir}| |\gray{proxy-roughness}| |\gray{--modules}| LoadData:proxy-roughness/LogData/data.npz PlotScatter:"dict(color='num_octaves',cmap='cet_bjy',grid=True)"
\end{lstlisting}

\begin{lstlisting}[frame=single, escapechar={|},
    caption={Full commands for generating Fig.~\ref{fig:roughness_and_slope_render_grid}: A set of terrain elements is saved and used in the second script. We loop over target-slope and target-roughness, and use these in \texttt{SetRoughness} and \texttt{SetSlope}.}
]
|\gray{python}| |\gray{generate\_terrain.py}| |\gray{--save-dir}| |\gray{slope\_grid}| --modules Octaves Save
|\gray{blender}| |\gray{--python}| |\gray{generate\_terrain.py}| |\gray{--}| |\gray{--save-dir}| |\gray{slope\_grid}| |\gray{--modules}| Loop:target_slope_deg=linspace[5,20,4] Loop:target_roughness=linspace[1.01,1.07,4] Load:slope_grid/Save Octaves:"dict(random_amp=0.0,random_sign=False,only_generate_weights=True)" Slope SetSlope Roughness SetRoughness WeightedSum Ground Camera:65 Render ClearScene ClearTerrain
\end{lstlisting}

\begin{lstlisting}[frame=single, escapechar={|},
    caption={Full commands for generating Fig.~\ref{fig:num_rocks_1}: Loop over \texttt{rock\_size} and \texttt{fraction} and generate \texttt{Rocks}. Collect data, and plot scatter of number of rocks.}
]
|\gray{python}| |\gray{generate\_terrain.py}| |\gray{--save-dir}| |\gray{rocksize\_fraction}| |\gray{--module}| Resolution:10 Loop:fraction=linspace[0.5,0.95,10] Loop:rock_size=linspace[0.5,4,15] Rocks FindRocks SurfaceStructure LogData
|\gray{python}| |\gray{generate\_terrain.py}| |\gray{--save-dir}|  |\gray{rocksize\_fraction}| |\gray{--module}| LoadData PlotScatter:"dict(color='fraction',cmap='cet_bjy',grid=True)"
\end{lstlisting}

\begin{lstlisting}[frame=single, escapechar={|},
    caption={Full command for generating Fig.~\ref{fig:num_rocks_grid}: Loop over \texttt{rock\_size} and \texttt{fraction} and generate \texttt{Rocks}. Find and plot rocks, and use as image texture in render.}
]
|\gray{blender}| |\gray{--python}| |\gray{generate\_terrain.py}| |\gray{--}| |\gray{--save-dir}| |\gray{rocksize\_fraction}| |\gray{--module}| Resolution:10 Loop:fraction=linspace[0.5,0.8,4] Loop:rock_size=linspace[1,4,4] Rocks FindRocks SurfaceStructure PlotObstacles ClearScene Ground ImageTexture Camera:45 Holdout Render --settings exportmode:True
\end{lstlisting}

\begin{lstlisting}[frame=single, escapechar={|},
    caption={Full commands for generating Fig.~\ref{fig:num_rocks_2}: Loop over \texttt{rock\_size} and plot scatter of height intervals.}
]
|\gray{python}| |\gray{generate\_terrain.py}| |\gray{--save-dir}| |\gray{height\_intervals}| |\gray{--module}| Resolution:10 Loop:rock_size=linspace[0.5,4,200] Rocks FindRocks SurfaceStructure LogData
|\gray{python}| |\gray{generate\_terrain.py}| |\gray{--save-dir}| |\gray{height\_intervals}| |\gray{--module}| LoadData PlotScatter:"dict(grid=True)"
\end{lstlisting}

\begin{lstlisting}[frame=single, escapechar={|},
    caption={Full commands for generating Fig.~\ref{fig:num_rocks_spatial}: A rock terrain, scaled by a \texttt{Basic:10} mask to create a spatial distribution of rocks.
The second command renders a plane with the mask as texture, and the two images have been combined.}
]
|\gray{blender}| |\gray{--python}| |\gray{generate\_terrain.py}| |\gray{--}| |\gray{--save-dir}| |\gray{spatial}| |\gray{--module}| Resolution:10 Basic:10 Scale:10 Clip Plot AsFactor Rocks Combine Scale FindRocks SurfaceStructure PlotObstacles Ground ImageTexture Camera:65 Render --settings exportmode:True
|\gray{blender}| |\gray{--python}| |\gray{generate\_terrain.py}| |\gray{--}| |\gray{--save-dir}| |\gray{spatial}| |\gray{--module}| Resolution:10 Plane Ground ImageTexture:spatial/Plot/terrain_temp_00000_0.png Camera:65 Holdout Render:mask.png
\end{lstlisting}

\begin{lstlisting}[frame=single, escapechar={|},
    caption={Full command for generating Fig.~\ref{fig:num_rocks_hexagon}. Show restrictions in rock positions without \texttt{random\_shift}.}
]
|\gray{python}| |\gray{generate\_terrain.py}| |\gray{--save-dir}| |\gray{hexagon}| |\gray{--module}| Loop:1000 Rocks:"dict(fraction=0.9,rock_size=4)" EndLoop Combine:Max Plot
\end{lstlisting}

\subsection{List of modules}
\label{subsec:list_of_modules}

\renewcommand{\arraystretch}{1.1} 
\setlength{\tabcolsep}{10pt}      

\begin{tabularx}{\textwidth}{>{\raggedright\arraybackslash}X
                            >{\raggedright\arraybackslash}X
                            >{\raggedright\arraybackslash}X
                            >{\raggedright\arraybackslash}X}

\underline{Generation Modules} &
\underline{Input/Output} &
\underline{Camera and Rendering} &
\underline{Terrain Analysis} \\

\texttt{Basic}, \texttt{Octaves}, \texttt{Rocks}, \texttt{Holes} &
\texttt{Save}, \texttt{Load} &
\texttt{Ground}, \texttt{Camera}, \texttt{ImageTexture}, \texttt{ColorMap}, \texttt{AddMeshObjects}, \texttt{Depth}, \texttt{Render}, \texttt{RenderSegmentation}, \texttt{ClearScene} &
\texttt{Roughness}, \texttt{Slope}, \texttt{SurfaceStructure}, \texttt{SetRoughness}, \texttt{SetSlope}, \texttt{FindRocks} \\[50pt]

\underline{Generation Functions} &
\underline{Modification and Math} &
\underline{Obstacles} &
\underline{Visualization} \\

\texttt{Function}, \texttt{Gaussian}, \texttt{Step}, \texttt{Donut}, \texttt{Plane}, \texttt{Sphere}, \texttt{Cube}, \texttt{SmoothStep}, \texttt{Sine} &
\texttt{Negate}, \texttt{Scale}, \texttt{Add}, \texttt{Absolute}, \texttt{Clip}, \texttt{Around}, \texttt{Smooth}, \texttt{AsProbability}, \texttt{AsLookupFor}, \texttt{AsFactor} &
\texttt{Random}, \texttt{PlotObstacles}, \texttt{SaveObstacles}, \texttt{LoadObstacles}, \texttt{RemoveDistantObstacles} &
\texttt{Plot}, \texttt{PlotRocks}, \texttt{PlotScatter}, \texttt{PlotHistogram}, \texttt{PlotLines} \\[40pt]

\underline{Compose and Combine} &
\underline{Basic Setup and Control} &
\underline{Sampling and Distribution} &
\underline{Terrain Configuration} \\

\texttt{Combine}, \texttt{WeightedSum}, \texttt{Stack} &
\texttt{Exit}, \texttt{Seed}, \texttt{Set}, \texttt{Print}, \texttt{DebugPlot} &
\texttt{SetDistribution}, \texttt{Sample}  &
\texttt{Extent}, \texttt{Size}, \texttt{Location}, \texttt{Resolution}, \texttt{GridSize}, \texttt{ClearTerrain} \\[10pt]

\underline{Data Processing} & \underline{Control Flow} & \\
\texttt{SaveData}, \texttt{LoadData}, \texttt{LogData} & \texttt{Loop}, \texttt{EndLoop}& \\

\end{tabularx}

\end{document}